\documentclass[aps,letterpaper,twocolumn,footinbib,floatfix,showpacs,hyperref,reprint]{revtex4}
\usepackage{amsmath}
\usepackage{amsfonts}
\usepackage{amssymb}
\usepackage{graphicx}
\usepackage{color}
\usepackage[caption=false]{subfig}
\usepackage{hyperref}
\usepackage[capitalize]{cleveref}

\usepackage{yypreamble}

\begin{document}

\title{Evolution of Coherence During Ramps Across the Mott-Superfluid Phase Boundary}

\author{Yariv Yanay}
\affiliation{Department of Physics, McGill University, Montreal, Canada H3A 2T8}
\author{Erich J. Mueller}
\affiliation{Laboratory of Atomic and Solid State Physics, Cornell University, Ithaca NY 14850}

\date{\today}

\begin{abstract}
We calculate how correlations in a Bose lattice gas grow during a finite speed ramp from the Mott to the Superfluid regime. We use an interacting doublon-holon model, applying a mean-field approach for implementing hard-core constraints between these degrees of freedom. Our solutions are valid in any dimension, and agree with experimental results and with DMRG calculations in one dimension. We find that the final energy density of the system drops quickly with increased ramp time for ramps shorter than one hopping time, $J\tau_{ramp}\lesssim 1$. For longer ramps, the final energy density depends only weakly on ramp speed. We calculate the effects of inelastic light scattering during such ramps.
\end{abstract}


\maketitle

\section{Introduction}

The dynamics of systems driven through a phase transition are a source of rich physics \cite{Sethna2006}. The phenomenology is particularly interesting in zero-temperature systems driven through a quantum phase transition \cite{Sondhi1997,Aoki2014}. In recent years, breakthrough experimental techniques in atomic physics have given us a direct probe of such transitions \cite{Orzel2001,Greiner2002,Stoferle2004,Hung2010,Chen2011}. In this paper, we model a bosonic lattice system driven from a Mott insulator state to into the superfluid regime. We introduce a novel mean-field theory, building on commonly used doublon-holon models \cite{Barmettler2012}. We calculate how correlations develop during a lattice ramp through the phase transition.

The phase diagram of bosonic lattice systems has been explored thoroughly \cite{Fisher1989,Sheshadri1993,Jaksch1998,Tasaki1998,Gurarie2009}. In the strongly interacting regime, at commensurate filling, lattice bosons form an incompressible Mott insulator. Conversely, for weak interactions the ground state is a superfluid Bose-Einstein condensate with long range order. When the system begins in a Mott insulator state and interactions are turned off, correlations grow as quasiparticles propagate across the system \cite{Cheneau2011,Natu2012}.

The Mott and superfluid phases can be approximated by distinct mean-field quasiparticle models. The excitations in the superfluid phase are well described by Bogoliubov quasiparticles made up of superpositions of particles and holes \cite{Griffin1996}. In the Mott insulator regime, on-site number fluctuations are small and the occupation of each site can be truncated to a small number of possibilities \cite{Barmettler2012}, the ``doublon-holon'' model. At strong coupling, the doublons and holons can be approximated as noninteracting bosons. These two descriptions are incompatible, making it challenge to model the dynamics across the phase boundary.

Previous work has produced partial understanding of this transition\cite{Polkovnikov2011,Dziarmaga2010}. Product state methods such as the Gutzwiller ansatz cannot calculate correlations \cite{Zwerger2003,Zakrzewski2005}. Other approaches have included calculations on small lattices \cite{Cucchietti2007}, field theory calculations for large particle density \cite{Amico1998,Polkovnikov2005,Navez2010,Tylutki2013} and various numerical techniques, which work well in one dimension but are otherwise more limited \cite{Kashurnikov2002,Kollath2007,Sau2012}. There has also been significant work on sudden quenches \cite{Polkovnikov2002,Tuchman2006} Here, we provide an analytical model that is particularly suitable for the small mean occupation numbers common in atomic experiments, provides access to coherence data, and is applicable in any number of dimensions.

\section{Model}
We perform our calculation within an approximate doublon-holon model. We restrict the state of each site $i$ to the subspace of occupations $\ket{i} \in \acom{\ket{\bar n + 1}, \ket{\bar n}, \ket{\bar n - 1}}$, where $\bar n$ is the median number of particles per site. The system can then be thought of in terms of a mean-occupation background and hard-core quasiparticle excitations of ``holons'' (an $\bar n - 1$ occupation) and ``doublons'' ($\bar n + 1$ occupation).
The annihilation operators at site $i$ for these quasiparticles are defined by 
$\hat d_{i}\ket{\bar n}_{i} = \hat d_{i}\ket{\bar n - 1}_{i} = \hat h_{i}\ket{\bar n + 1}_{i} = \hat h_{i}\ket{\bar n}_{i} = 0$,  ${\hat d_{i}\ket{\bar n + 1}_{i} = \hat h_{i}\ket{\bar n - 1}_{i} = \ket{\bar n }_{i}}$.

Under this approximation, the Hamiltonian is
\begin{equation} \begin{split}
\hat H  =  & \sum_{k}\br{ \half[U] + J \sqrt{\tilde n^{2} +\tfrac{1}{4}}\gve_{k}}\p{\hat d_{k}\dg\hat d_{k} + \hat h_{k}\dg \hat h_{k}}
\\ + &\half[J] \gve_{k} \p{\hat d_{k}\dg\hat d_{k} - \hat h_{k}\dg \hat h_{k}} +  J\tilde n\gve_{k}\p{\hat d_{k}\hat h_{-k} + \hat h_{-k}\dg \hat d_{k}\dg}.
\label{eq:HDH}
\end{split} \end{equation}
Here, $\hat d_{k} = \tfrac{1}{\sqrt{N_{s}}}\sum_{i}e^{i\vk\cdot\vec r_{i}}\hat d_{i}$, summing over all sites $i$, and similar for $\hat h_{k}$, while $\gve_{k} = -2\sum_{\Delta}\cos\p{\vk\cdot\vec \Delta}$, summing over lattice basis vectors, $\vec\Delta = \Delta\hat x, \Delta\hat y, \Delta\hat z$ in three dimensions, or a subset of those in lower dimensions. These represent a cubic lattice with lattice constant $\Delta$. $U$ and $J$ are the interaction and hopping strength, respectively, and $\tilde n = \sqrt{\bar n\p{\bar n + 1}}$. $N_{s}$ is the number of sites in the lattice.

The doublon-holon model is an approximation for the single-band Bose-Hubbard model \cite{Hubbard1963,Fisher1989}. It is most accurate in the low-temperature, strongly-interacting limit, as the energy of a state increases quadratically with the deviation from the mean particle number. However, for low occupation numbers $\bar n$, it can be a good approximation in the weakly-interacting limit as well. In a noninteracting superfluid gas with $\bar n = 1$, the probability of finding more than two particles on a given site is less than 10\%. We do all our calculations in this regime, taking, $\avg{\hat n_{i}} = \bar n + \avg{\hat d_{i}\dg\hat d_{i}} - \avg{\hat h_{i}\dg\hat h_{i}} = \bar n = 1$.

We calculate the time evolution of the two-point correlation functions, ${\avg{\hat d_{k}\dg\hat d_{k}} \approx \avg{\hat h_{k}\dg\hat h_{k}}}$ and $\avg{\hat d_{k}\hat h_{-k}}$ using the Heisenberg equation, $\frac{d}{dt}\avg{\hat X} = i\br{\hat H, \hat X}$. The hard-core constraints for $\hat d_{i},\hat h_{i}$ imply nontrivial commutation relations, and the resulting equations of motion involve four-point correlation functions such as 
\begin{equation} \begin{split}
C_{k,p,q} = \avg{\hat d_{p}\dg\hat h_{-p-q}\dg\hat h_{-k-q}\hat d_{k}}.
\end{split} \end{equation}
We can characterize $C_{k,p,q}$ by writing it in the form
\begin{equation} \begin{split}
C_{k,p,q} & = \gd_{p,k}\avg{\hat h_{-k-q}\dg\hat h_{-k-q}}\avg{\hat d_{k}\dg\hat d_{k}}
\\ & + \gd_{q,0}\avg{\hat d_{p}\dg\hat h_{-p}\dg}\avg{\hat h_{-k}\hat d_{k}} - \tfrac{\ga_{k,p,q}}{N_{s}}\avg{\hat d_{k}\dg\hat d_{k}}.
\end{split} \end{equation}
This equation defines the function $\ga_{p,k,q}$. We make a mean-field approximation, taking ${\ga_{p,k,q} \approx \tfrac{1}{n_{d}}\avg{\hat h_{-p-q}\dg\hat h_{-p-q}}\avg{\hat d_{p}\dg\hat d_{p}}}$, where $n^{d} = \tfrac{1}{N_{s}}\sum_{k}\avg{\hat d_{k}\dg\hat d_{k}}$ is the doublon density. This approximation enforces the hard-core constraint $\sum_{k}C_{k,p,q} = 0$, and becomes exact in the deep Mott regime. We make similar approximation for the other four-point correlation functions, as described in detail in \cref{sec:deriv}.

We arrive at a closed set of non-linear, coupled differential equations that we numerically integrate to find all quasiparticle two-point correlation functions at any time. From these we can easily extract the correlation functions for real particles, $\avg{\hat a_{i}\dg\hat a_{j}}$ and $\avg{\hat a_{k}\dg\hat a_{k}}$.

\begin{figure}[t] 
   \centering
   \includegraphics[width=\columnwidth]{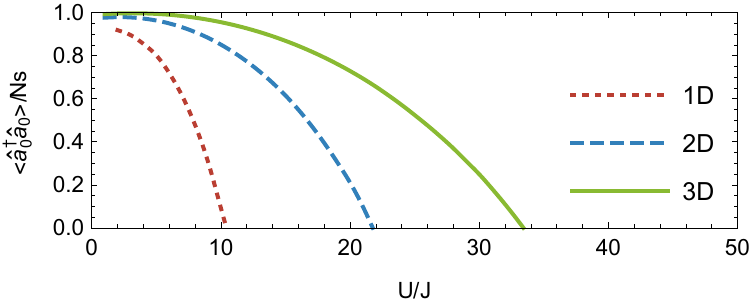} 
   
   \includegraphics[width=\columnwidth]{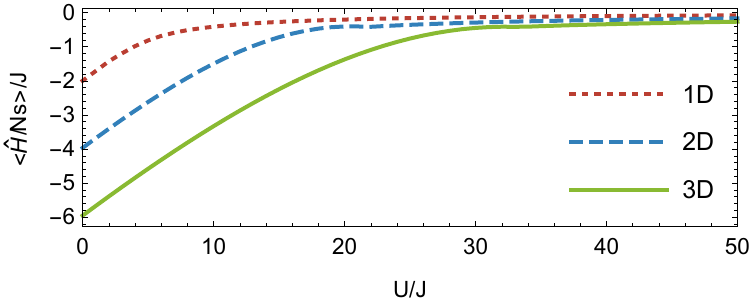} 
   \caption{(Color Online) Equilibrium properties of the hard-core doublon-holon model discussed in the text for a cubic lattice with mean filling $\bar n = 1$. 
Shown  as a function of the interaction strength $U/J$, above: the equilibrium condensate fraction, below: the average energy per particle in the ground state.}
   \label{fig:eqphase}
\end{figure}

\section{Equilibrium State}

We find the equilibrium state under this model by minimizing the expectation value $\avg{\hat H}$ of the Hamiltonian of \cref{eq:HDH} while requiring $\frac{d}{dt}\avg{\hat d_{k}\dg\hat d_{k}} = \frac{d}{dt}\avg{\hat d_{k}\hat h_{-k}} = 0$. As seen in \cref{fig:eqphase}, we find a phase transition at a critical value of $U_{c}/J = 10.4,21.8,33.4$ in one-, two- and three-dimensions. These are similar to the standard mean-field values of $U_{c}/J = 11.6, 23.2, 34.8$ \cite{Bloch2008,Sheshadri1993} and somewhat higher than numerically calculated values $U_{c}/J = 3.6,16.9,29.3$ \cite{Freericks1994,Monien1998,Elstner1999,Capogrosso-Sansone2007,Capogrosso-Sansone2007a,Teichmann2009}.

\begin{figure}[t] 
   \centering
   \includegraphics[width=\columnwidth]{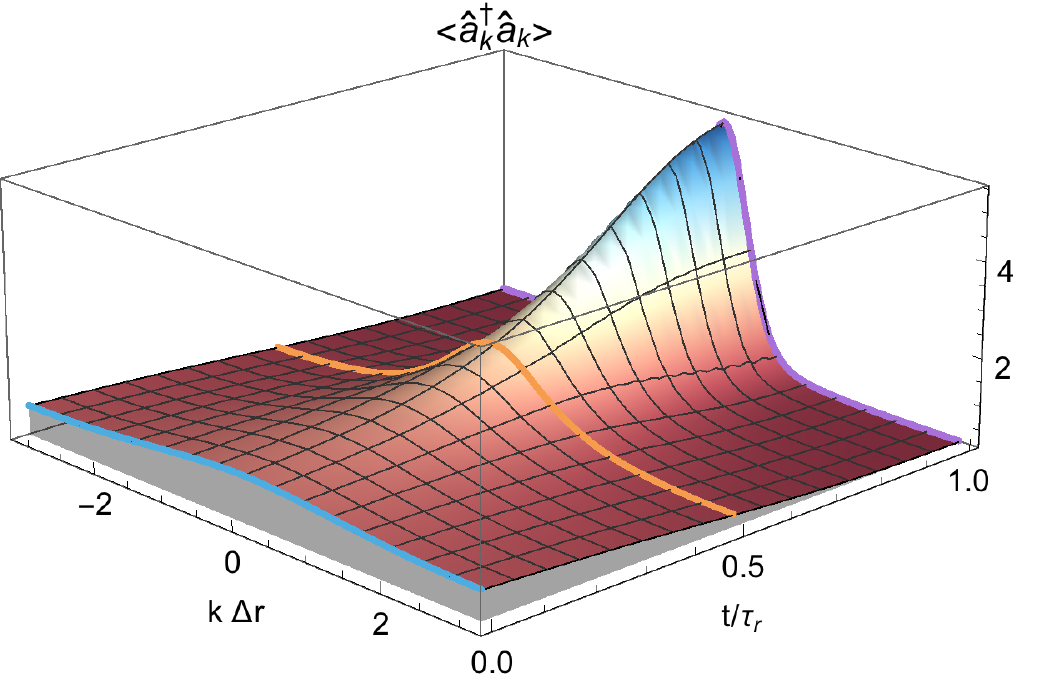}
   \caption{(Color Online) Evolution of the momentum density distribution function $\avg{\hat a_{k}\dg\hat a_{k}}$ as the interaction strength is slowly ramped down, $U = U_{i}\p{U_{f}/U_{i}}^{t/\tau_{r}}$ in a one-dimensional lattice. Here $U_{i} = 47J, U_{f} = 2J, J\tau_{r} = 2$. 
}
   \label{fig:nkex}
\end{figure}

\section{Interaction Ramps}

We use the model above to explore the behavior of a gas subject to a non-adiabatic ramp of the interaction through the phase transition. We perform an interaction ramp of the form 
\begin{equation} \begin{split}
U = U_{i}\p{U_{f}/U_{i}}^{t/\tau_{r}},
\end{split} \end{equation}
where the ground state of the system is a Mott insulator for $U = U_{i}$ and superfluid for $U = U_{f}$. The time scale $\tau_{r}$ sets the speed of the ramp. This form approximates the relation $U/J$ in an optical lattice experiment if the scattering length is fixed and the lattice depth is ramped down \cite{Jaksch1998}.

We initialize the system in the ground state at the initial lattice depth, in the Mott regime, and perform a finite-element time integration of the evolution equations as the interaction strength is reduced. We calculate the momentum space density throughout this evolution for various values of $\tau_{r}$. \Cref{fig:nkex} shows the behavior for a typical ramp, with $J\tau_{r} = 2$. We have full access to all two-point observables at any time along the ramp.

We first characterize the behavior of the system at the end of the ramp. We define an effective correlation length, $\xi$, by comparing correlations in the system to the form $\avg{\hat a_{i}\dg\hat a_{j}} = \bar n e^{-\abs{\vec r_{j} - \vec r_{i}}/\xi}$.  We calculate $\xi$ by fitting to the width of the momentum distribution, as defined by the first moment, yielding
\begin{equation} \begin{split}
\frac{\xi}{\Delta} = -1/\log \br{\tfrac{1}{N_{s}}\sum_{k}\frac{\gve_{k}}{\gve_{0}}\avg{\hat a_{k}\dg\hat a_{k}}}.
\label{eq:xi}
\end{split} \end{equation}
Though it is infinite for an equilibrium superfluid system, $\xi$ remains finite at any finite time for a system that is not initially superfluid \cite{Cheneau2011}.

\begin{figure}[h] 
   \centering
   \includegraphics[width=\columnwidth]{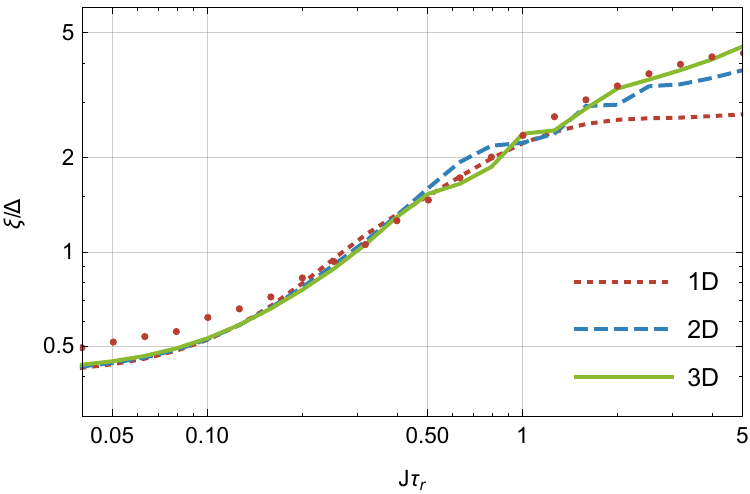} 
   \caption{(Color Online) Effective correlation length $\xi$ (see \cref{eq:xi}), normalized by the lattice constant $\Delta$, at the end of a ramp of the interaction strength of the form ${U = U_{i}\p{U_{f}/U_{u}}^{t/\tau_{r}}}$. Here $U_{i} = 47J, U_{f} = 2J$.
   The red dots are the result of an exact diagonalization calculation for an 11-site one-dimensional lattice.
}
   \label{fig:endoframp}
\end{figure}

\Cref{fig:endoframp} shows the effective correlation length at the end of the ramp for varying ramp times.  In one dimension our calculation agrees well with the result of an exact diagonalization of a small lattice. The discrepancy is consistent with the finite-size effects in the exact diagonalization.  Our results also agree with the experimental results of \cite{Braun2014}. For slow ramps we see that the correlation length is somewhat smaller in lower dimensions.

\section{Final Energy Density}

After the ramp has ended, the system continues to evolve, and the correlation length continues to grow. However, the energy of the system is now conserved. At long times after the ramp we expect the state of the system to resemble a thermal state at a temperature determined by the energy density $\mathcal U = \tfrac{1}{N_{s}}\avg{\hat H} - \avg{\hat H}_{gs}$, where $\avg{\hat H}_{gs}$ is the energy of the new ground state of the system. 

\begin{figure}[h] 
   \centering
   \includegraphics[width=\columnwidth]{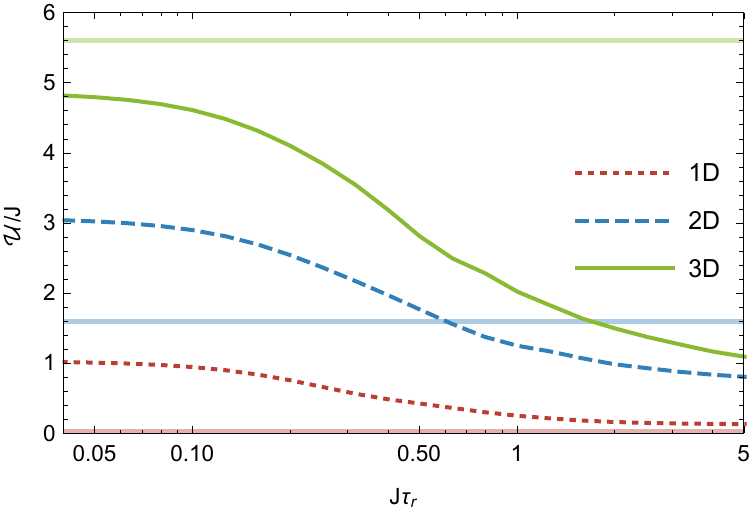} 
   \caption{(Color Online) The energy density $\mathcal U$ following an interaction ramp of the form $U = U_{i}\p{U_{f}/U_{i}}^{t/\tau_{r}}$. Here ${U_{i} = 47J}$, $U_{f} = 2J$.
     Horizontal lines show the energy density, $\mathcal U_{c}/J = 0, 2.1, 5.1$, at the superfluid critical temperature $T_{c}/J = 0, 1.7, 5.9$, for $U = U_{f} = 2J$, in one, two and three dimensions \cite{Capogrosso-Sansone2007a,Capogrosso-Sansone2007}.}
   \label{fig:Eendoframp}
\end{figure}

We plot $\mathcal U$ as a function of the ramp time $\tau_{r}$ in \cref{fig:Eendoframp}. For ramp times much shorter than the hopping time scale, $J\tau_{r}\lesssim 0.2$, the final energy density varies slowly with $\tau_{r}$. Such ramps are indistinguishable from instantaneous quenches, and the final state of the system, if allowed to equilibrate, would be similar for any $\tau_{r}$ in this regime. For $J\tau_{r}\gtrsim 0.2$, the system's energy depends more strongly on the length of the ramp.

\Cref{fig:Eendoframp} also shows the critical energy density $\mathcal U_{c}$ corresponding to the energy density of of a Bose lattice gas with $U=U_{f}$ at the critical temperature of the superfluid-normal gas phase transition \cite{Capogrosso-Sansone2007a,Capogrosso-Sansone2007}. We expect a gas with $\mathcal U>\mathcal U_{c}$ to equilibrate to a normal-gas state with finite correlation length $\xi$, while a gas at $\mathcal U<\mathcal U_{c}$ would equilibrate to a superfluid state with long-range order. In two dimensions, we expect short ramps, $J\tau_{r}\lesssim 0.6$, to lead to a normal state, while longer ramps lead to a superfluid gas. In three dimensions, the energy density is always below $\mathcal U_{c}$, even for an instantaneous quench. In one dimension there is no condensed phase.

\section{Decoherence}

In an ideal, closed, quantum system, all evolution is unitary. The final energy of the system rises monotonously with the rate of the ramp in such systems. Conversely, any real system suffers from heating, atom loss and other impacts from the environment. As a result, experimental dynamic systems always face a competition between the system's reaction time and external processes.

The physics of such decoherence has been explored in detail \cite{Schachenmayer2014,Cai2013,Poletti2013,Pichler2013,Poletti2012,Pichler2010}. Here, we return to a mechanism we have previously used to described the effect of density measurement by light scattering \cite{Yanay2014}. The same formalism describes inelastic light scattering, where an external photon scatters off of a trapped atom. This is one of the major sources of decoherence in atomic experiments.

As in \cite{Yanay2014}, we neglect out of band effects, which cause particle loss. We focus on in-band scattering, which would directly decrease the coherence of the remaining gas and reduce the correlation length measured above. In an ensemble description, this leads to a nonunitary evolution term of the form
\begin{equation} \begin{split}
\frac{d}{dt}\avg{\hat d_{k}\dg\hat d_{k}} = -&i\avg{\br{\hat d_{k}\dg\hat d_{k},\hat H}} - \gamma\p{\avg{\hat d_{k}\dg\hat d_{k}} - n_{d}}
\\ \frac{d}{dt}\avg{\hat d_{k}\hat h_{-k}} & = -i\avg{\br{\hat d_{k}\hat h_{-k},\hat H}} - \gamma\avg{\hat d_{k}\hat h_{k}}
\label{eq:ddecoh}
\end{split} \end{equation}
where $\gamma$ is proportional to the frequency of light scattering per site.

\begin{figure}[htbp] 
   \centering
   \includegraphics[width=\columnwidth]{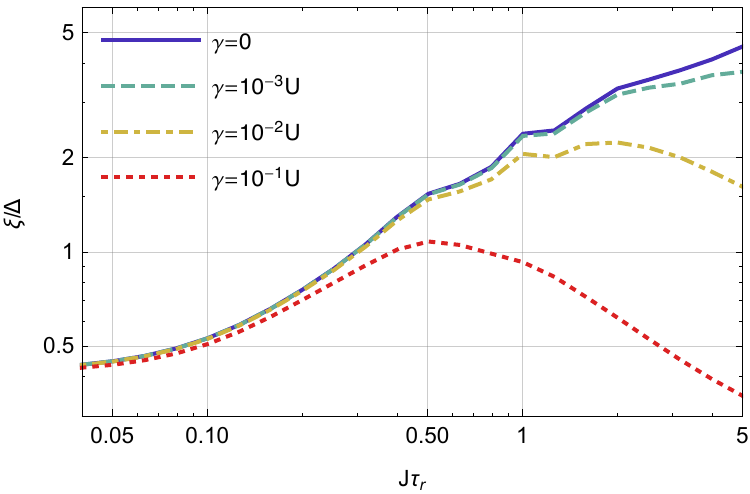} 
 \caption{(Color Online) Effective correlation length $\xi$, at the end of a ramp of the interaction strength,
 in a system coupled to the environment in the form shown in \cref{eq:ddecoh}. Here $U_{f} = 47J, U_{i} = 2J$, in a three-dimensional cubic lattice.
   In an optical lattice setup, the rate of inelastic light scattering events changes with lattice depth similarly to the interaction strength \cite{Jaksch1998,Pichler2010}.}
   \label{fig:decoh}
\end{figure}

We calculate the effect of this decoherence on the behavior of the correlation length $\xi$, as shown in \cref{fig:decoh}. As expected, no effect is seen at time scales shorter than $1/\gamma$, but at longer time scales, inelastic processes cause the correlation length to decay. The overall effect is similar to experimental observations in \cite{Braun2014}.

\section{Outlook}

The physics of ultracold atomic systems involves multiple energy scales. In driven experimental systems, these include the relaxation time of the system, the driving time scale and the rate of decoherence imposed by interaction with the environment. Here, we have quantified the effect of the quench rate in Bose-Hubbard systems crossing the phase boundary. We find that there are two regimes. For sweeps which are much shorter than the typical hopping time, $J\tau_{r} \lesssim 0.2$, the ramp time has no effect on the final state and the ramp is indistinguishable from an instantaneous quench. For longer ramps, the final energy density of the state and therefore its correlations at equilibrium, depend on the length of the ramp. In two dimensions, shorter ramps lead to a normal gas state, while longer ramps lead to a superfluid state. We have also demonstrated that inelastic light scattering can be quite destructive on longer time scale, underscoring the usefulness of shorter experimental runs.

\section*{Acknowledgements}
We acknowledge support from the ARO-MURI Non-equilibrium Many-body Dynamics grant (63834-PH-MUR).

\appendix

\newpage
\widetext
\section{Detailed Derivation of the Approximations in the Hard-Core Doublon-Holon Model\label{sec:deriv}}

\subsection{Underlying Model}

We perform our calculation within an approximate ``doublon-holon'' model. The state of each site $i$ can be given in terms of a spinor in the allowed occupation states
$\ket{\bar n + 1}_{i}, \ket{\bar n}_{i}, \ket{\bar n - 1}_{i}$ where $\bar n$ is the median number of particles per site. We define the quasiparticle annihilation operators, $\hat d_{i} = \ket{\bar n}\bra{\bar n + 1}_{i}, {\hat h_{i} = \ket{\bar n}\bra{\bar n - 1}_{i}}.$

Under this approximation, the Hamiltonian is
\begin{equation} \begin{split}
\hat H & = -J\sum_{\avg{i,j}}\p{\p{\bar n + 1}\hat d_{i}\dg\hat d_{j} + \bar n \hat h_{i}\dg\hat h_{j} + \sqrt{\bar n\p{\bar n + 1}}\p{\hat d_{i}\hat h_{j} + \hat d_{j}\dg\hat h_{i}\dg} + \hc}
+ \frac{U}{2}\sum_{i}\p{\hat d_{i}\dg\hat d_{i} + \hat h_{i}\dg\hat h_{i}}
\\ & =  \sum_{k}\br{\p{ \half[U] + J \sqrt{\tilde n^{2} +\tfrac{1}{4}}\gve_{k}}\p{\hat d_{k}\dg\hat d_{k} + \hat h_{k}\dg \hat h_{k}}
+ \half[J] \gve_{k} \p{\hat d_{k}\dg\hat d_{k} - \hat h_{k}\dg \hat h_{k}} +  J\tilde n\gve_{k}\p{\hat d_{k}\hat h_{-k} + \hat h_{-k}\dg \hat d_{k}\dg}}
\label{eq:supHDH}
\end{split} \end{equation}
Here, 
\begin{equation} \begin{split}
\hat d_{k} = \tfrac{1}{\sqrt{N_{s}}}\sum_{i}e^{i\vk\cdot\vec r_{i}}\hat d_{i}, \qquad \hat h_{k} = \tfrac{1}{\sqrt{N_{s}}}\sum_{i}e^{i\vk\cdot\vec r_{i}}\hat h_{i},
\end{split} \end{equation}
summing over all sites $i$, and 
\begin{equation} \begin{split}
\gve_{k} = -2\sum_{\Delta}\cos\p{\vk\cdot\vec \Delta}
\end{split} \end{equation}
summing over lattice vectors. We perform our calculation on a cubic lattice with lattice spacing $\Delta$, $\vec\Delta = \Delta\hat x, \Delta\hat y, \Delta\hat z$ in three dimensions or a subset of those in lower dimensions. $U$ and $J$ are the interaction and hopping strength, respectively, and $\tilde n = \sqrt{\bar n\p{\bar n + 1}}$. $N_{s}$ is the number of sites in the lattice.

We do all our calculations for a density of one particle per site, $\avg{\hat n_{i}} = \bar n + \avg{\hat d_{i}\dg\hat d_{i}} - \avg{\hat h_{i}\dg\hat h_{i}} = \bar n = 1$.

The hard-core constraints on the operators $\hat d$, $\hat h$ translate into non-trivial commutation relations,
\begin{equation} \begin{split}
\br{\hat d_{k}, \hat d_{q}\dg} & = \gd_{k,q} - 2\hat n^{d}_{q-k} - \hat n^{h}_{q-k}, \quad \br{\hat d_{k}\dg, \hat h_{q}} = \hat\nu_{q-k}
\\ \br{\hat h_{k}, \hat h_{q}\dg} & = \gd_{k,q} - \hat n^{d}_{q-k} - 2\hat n^{h}_{q-k}, \quad \br{\hat h_{k}\dg, \hat d_{q}} = \hat\nu_{k-q}\dg
\\ & \br{\hat d_{k}, \hat d_{q}} = \br{\hat h_{k}, \hat h_{q}} = \br{\hat d_{k}, \hat h_{q}}  = 0,
\label{eq:com}
\end{split} \end{equation}
where we define the quasiparticle density operators
\begin{equation} \begin{split}
\hat n^{d}_{k} = \tfrac{1}{N_{s}}\sum_{i}e^{-i\vk\cdot r_{i}}\hat d_{i}\dg\hat d_{i}, &
\quad 
\hat n^{h}_{k} = \tfrac{1}{N_{s}}\sum_{i}e^{-i\vk\cdot r_{i}}\hat h_{i}\dg\hat h_{i},
\quad 
\hat \nu_{k}\dg = \tfrac{1}{N_{s}}\sum_{i}e^{-i\vk\cdot r_{i}}\hat h_{i}\dg\hat d_{i}.
\label{eq:densops}
\end{split} \end{equation}

We write $\hat n_{d,h} \equiv \hat n^{d,h}_{0}$, the density of doublons and holons, respectively. In the Mott equilibrium limit, the operators in \cref{eq:densops} can be neglected and the quasiparticles can be treated as noninteracting bosons. This is not true in the superfluid regime.

\subsection{Equations of Motion}
Equations of motion can be derived from the Hamiltonian, \cref{eq:supHDH}, via the Heisenberg equation,
\begin{equation} \begin{split}
\frac{d}{dt}\avg{\hat X} = i\br{\hat X, \hat H}.
\end{split} \end{equation}

 We focus on the two-point observables, 
\begin{equation} \begin{split}
\frac{d}{dt}\avg{\hat d_{k}\dg\hat d_{k}} & = i J \tilde n\gve_{k}\p{\avg{\hat d_{k}\hat h_{-k}} - \avg{\hat d_{k}\dg\hat h_{-k}\dg}}
\\ & \quad -iJ\tilde n \sum_{q}\gve_{q}\br{\begin{array}{c}
\p{\sqrt{1 + \tfrac{1}{4\tilde n^{2}}} + \tfrac{1}{2\tilde n}}\avg{\hat d_{q}\dg\p{2\hat n^{d}_{k-q} + \hat n^{h}_{k-q}}\hat d_{k}}
\\  + \p{\sqrt{1 + \tfrac{1}{4\tilde n^{2}}} - \tfrac{1}{2\tilde n}} \avg{\hat h_{-q}\dg\hat\nu_{-k-q}\hat d_{k}}
\\ + \avg{\p{2\hat n^{d}_{k-q} + \hat n^{h}_{k-q}}\hat h_{-q}\hat d_{k}} + \avg{\hat \nu_{-q-k}\hat d_{q}\hat d_{k}}
\end{array} - \hc}
\\ \frac{d}{dt}\avg{\hat h_{-k}\dg\hat h_{-k}} & = i J \tilde n\gve_{k}\p{\avg{\hat d_{k}\hat h_{-k}} - \avg{\hat d_{k}\dg\hat h_{-k}\dg}}
\\ & \quad  -iJ\tilde n \sum_{q}\gve_{q}\br{\begin{array}{c}
\p{\sqrt{1 + \tfrac{1}{4\tilde n^{2}}} - \tfrac{1}{2\tilde n}}\avg{\hat h_{-q}\dg\p{2\hat n^{h}_{q-k} + \hat n^{d}_{q-k}}\hat h_{-k}}
\\  + \p{\sqrt{1 + \tfrac{1}{4\tilde n^{2}}} + \tfrac{1}{2\tilde n}} \avg{\hat d_{q}\dg\hat\nu_{-q-k}\dg\hat h_{-k}}
\\ + \avg{\p{2\hat n^{h}_{q-k} + \hat n^{d}_{q-k}}\hat d_{q}\hat h_{-k}} + \avg{\hat \nu_{-q-k}\dg\hat h_{-q}\hat h_{-k}}
\end{array} - \hc}
\\ \frac{d}{dt}\avg{\hat d_{k}\hat h_{-k}} & = -i U\avg{\hat d_{k}\hat h_{-k}} - iJ\tilde n\gve_{k}\p{1  - 3n_{d} - 3n_{h}}
\\ & \quad -iJ\tilde n \sum_{q}\gve_{q}\p{\gd_{k,q} - \tfrac{1}{N_{s}}}\br{2\sqrt{1 + \tfrac{1}{4\tilde n^{2}}}\avg{\hat d_{q}\hat h_{-q}}
 + \avg{\hat d_{q}\dg\hat d_{q}} + \avg{\hat h_{-q}\dg\hat h_{-q}}}
\\ & \quad +i J \tilde n \sum_{q}\gve_{q}\bmat{\p{\sqrt{1 + \tfrac{1}{4\tilde n^{2}}} + \tfrac{1}{2\tilde n}}
\mat{\avg{\hat h_{-q}\p{ 2\hat n^{d}_{k-q} + \hat n^{h}_{k-q}}\hat d_{k}} + \avg{\hat d_{q}\hat \nu_{-q-k}\hat d_{k}}}
\\ + \p{\sqrt{1 + \tfrac{1}{4\tilde n^{2}}} - \tfrac{1}{2\tilde n}}\mat{\avg{\hat d_{q}\p{ \hat n^{d}_{q-k} + 2\hat n^{h}_{q-k}}\hat h_{-k}} + \avg{\hat h_{-q}\hat \nu_{-q-k}\dg\hat h_{-k}}}
\\ + \avg{\hat h_{-q}\dg\p{2\hat n^{d}_{q-k} + \hat n^{h}_{q-k}}\hat h_{-k} } + \avg{\hat d_{q}\dg\hat \nu_{-q-k}\dg\hat h_{-k}}
\\ + \avg{\hat d_{q}\dg\p{\hat n^{d}_{k-q} + 2\hat n^{h}_{k-q}}\hat d_{k}} + \avg{\hat h_{-q}\dg \hat \nu_{-q-k} \hat d_{k}}
\\ - \p{2\hat n^{d}_{q-k} + \hat n^{h}_{q-k}}\p{\hat n^{d}_{k-q} + 2\hat n^{h}_{k-q}} - \half \p{\hat\nu_{-q-k}\dg\hat\nu_{-q-k} + \hat\nu_{-q-k}\hat\nu_{-q-k}\dg}
},
\label{eq:eomfull}
\end{split} \end{equation}
Here $\hc$ stands for the Hermitian conjugate. 

\subsection{Hard-Core Coherent Approximation}

To perform the time evolution, we must make approximations for the quartic terms, such as 
\begin{equation} \begin{split}
\mathcal C^{1}_{k} = \sum_{q}\gve_{q}\avg{\hat d_{q}\dg\hat n^{h}_{k-q}\hat d_{k}} = \tfrac{1}{N_{s}}\sum_{p,q}\gve_{q}\avg{\hat d_{q}\dg\hat h_{-p-q}\dg\hat h_{-p-k}\hat d_{k}}
\equiv \tfrac{1}{N_{s}}\sum_{p,q}\gve_{q}C_{k,q,p}.
\end{split} \end{equation}
These terms can be written out as
\begin{equation} \begin{split}
\mathcal C^{1}_{k} & = \tfrac{1}{N_{s}}\sum_{p}\gve_{k}\avg{\hat d_{k}\dg\hat h_{-k-p}\dg\hat h_{-k-p}\hat d_{k}} + \tfrac{1}{N_{s}}\sum_{q}\gve_{q}\avg{\hat d_{q}\dg\hat h_{-q}\dg\hat h_{-k}\hat d_{k}}
\\ & \quad - \tfrac{1}{N_{s}}\gve_{k}\avg{\hat d_{k}\dg\hat h_{-k}\dg\hat h_{-k}\hat d_{k}}
 + \tfrac{1}{N_{s}}\sum_{\substack{p,q\\q\ne k,p\ne 0}}\gve_{q}\avg{\hat d_{q}\dg\hat h_{-p-q}\dg\hat h_{-k-q}\hat d_{k}}.
 \label{eq:C1}
\end{split} \end{equation}
The first two sums on the right hand side add up coherently, and we expect them to dominate. The third term is inversely proportional to the system size, and is therefore negligible. For bosonic operators, one may expect the final sum to add up incoherently, as in the Hartree-Fock-Bogoliubov approximation \cite{Griffin1996}, suggesting the form
\begin{equation} \begin{split}
\mathcal C^{1}_{k} \approx \tilde{\mathcal C}^{1}_{k}= \p{\tfrac{1}{N_{s}}\sum_{p}\avg{\hat h_{-k-p}\dg\hat h_{-k-p}}}\gve_{k}\avg{\hat d_{k}\dg\hat d_{k}} + \p{\tfrac{1}{N_{s}}\sum_{q}\gve_{q}\avg{\hat d_{q}\dg\hat h_{-q}\dg}}\avg{\hat h_{-k}\hat d_{k}},
\end{split} \end{equation}
This intuition fails in the hard-core case. This can be seen by summing over the momenta,
\begin{equation} \begin{split}
\sum_{k}\mathcal C^{1}_{k} & = \tfrac{1}{N_{s}}\sum_{p,q,k}\gve_{q}\avg{\hat d_{q}\dg\hat h_{-p-q}\dg\hat h_{-p-k}\hat d_{k}}
=  \tfrac{1}{N_{s}}\sum_{p,q,i}e^{ipr_{i}}\gve_{q}\avg{\hat d_{q}\dg\hat h_{-p-q}\dg\hat h_{i}\hat d_{i}} = 0
\\ & \sum_{k}\tilde{\mathcal C}^{1}_{k} = \p{\tfrac{1}{N_{s}}\sum_{p}\avg{\hat h_{-k-p}\dg\hat h_{-k-p}}}\p{\sum_{k}\gve_{k}\avg{\hat d_{k}\dg\hat d_{k}}} \ne 0.
\end{split} \end{equation}
To account for the hard core constraints, we formally write
\begin{equation} \begin{split}
C^{1}_{k,p,q} = \avg{\hat d_{q}\dg\hat h_{-p-q}\dg\hat h_{-p-k}\hat d_{k}} = \gd_{k,q}\avg{\hat h_{-p-q}\dg\hat h_{-p-k}}{\hat d_{k}\dg\hat d_{k}} + \gd_{p,0}\avg{\hat d_{q}\dg\hat h_{-q}\dg}\avg{\hat h_{-k}\hat d_{k}} - \tfrac{\ga_{k,p,q}}{N_{s}}\avg{\hat d_{k}\dg\hat d_{k}},
\end{split} \end{equation}
where this equation defines $\ga_{k,p,q}$. We approximate this function with the hard core constraint in mind,
\begin{equation} \begin{split}
\ga_{k,p,q} \approx \frac{1}{\avg{\hat n^{d}}}\avg{\hat h_{-p-q}\dg\hat h_{-p-q}}\avg{\hat d_{q}\dg\hat d_{q}},
\end{split} \end{equation}
so that $\sum_{k}C^{1}_{k,p,q} = 0$. We then find
\begin{equation} \begin{split}
\mathcal C^{1}_{k} = \tfrac{1}{N_{s}}\sum_{p,q}C^{1}_{k,p,q} \approx \avg{\hat n^{h}}\mat{\gve_{k} - \gve_{0}\frac{\xi_{d}}{n_{d}}}\avg{\hat d_{k}\dg\hat d_{k}} + \gve_{0}\eta^{*}\avg{\hat d_{k}\hat h_{-k}}
\end{split} \end{equation}
where
\begin{equation} \begin{split}
n_{d} = \avg{\hat n^{d}} \qquad \xi_{d} = \tfrac{1}{N_{s}}\sum_{k}\tfrac{\gve_{k}}{\gve_{0}}\avg{\hat d_{k}\dg\hat d_{k}}
 \qquad  \eta = \tfrac{1}{N_{s}}\sum_{k}\tfrac{\gve_{k}}{\gve_{0}}\avg{\hat d_{k}\hat h_{-k}}.
\end{split} \end{equation}

We make similar approximations for the other terms,
\begin{equation} \begin{split}
\avg{\hat d_{q}\dg\hat d_{p-q}\dg\hat d_{p-k}\hat d_{k}} & \approx \gd_{q,k}\avg{\hat d_{p-k}\dg\hat d_{p-k}}\avg{\hat d_{k}\dg\hat d_{k}} + \gd_{p-q,k}\avg{\hat d_{q}\dg\hat d_{q}}\avg{\hat d_{k}\dg\hat d_{k}} - \tfrac{2}{N_{s}}\tfrac{1}{n_{d}}\avg{\hat d_{q}\dg\hat d_{q}}\avg{\hat d_{p-q}\dg\hat d_{p-q}}\avg{\hat d_{k}\dg\hat d_{k}}
\\ \avg{\hat h_{-q}\dg\hat d_{p+q}\dg\hat h_{p-k}\hat d_{k}} & \approx \gd_{p,0}\avg{\hat h_{-q}\dg\hat d_{q}\dg}\avg{\hat d_{k}\hat h_{-k}} + \gd_{p+q,k}\avg{\hat h_{-q}\dg\hat h_{-q}}\avg{\hat d_{k}\dg\hat d_{k}} - \tfrac{1}{N_{s}}\tfrac{1}{n_{d}}\avg{\hat h_{-q}\dg\hat h_{q}}\avg{\hat d_{p+q}\dg\hat d_{p+q}}\avg{\hat d_{k}\dg\hat d_{k}}
\\ \avg{\hat h_{-q}\hat h_{-p-q}\dg\hat h_{-p-k}\hat d_{k}} & \approx \\ \gd_{q,k}\avg{\hat h_{-p-k}\dg\hat h_{-p-k}}&\avg{\hat d_{k}\hat h_{-k}} + \gd_{p,0}\avg{\hat h_{-q}\hat h_{-q}\dg}\avg{\hat d_{k}\hat h_{-k}} - \tfrac{1}{N_{s}}\tfrac{1}{n_{d}}\avg{\hat h_{-p-q}\dg\hat h_{-p-q}}\avg{\hat d_{q}\hat h_{-q}}\avg{\hat d_{k}\dg\hat d_{k}}
\\ \avg{\hat h_{-q}\hat d_{p-q}\dg\hat d_{p-k}\hat d_{k}} & \approx \\ \gd_{q,k}\avg{\hat d_{p-k}\dg\hat d_{p-k}}&\avg{\hat d_{k}\hat h_{-k}} + \gd_{p-q,k}\avg{\hat d_{q}\hat h_{-q}}\avg{\hat d_{k}\dg\hat d_{k}} - \tfrac{2}{N_{s}}\tfrac{1}{n_{d}}\avg{\hat d_{q}\hat h_{-q}}\avg{\hat d_{p-q}\dg\hat d_{p-q}}\avg{\hat d_{k}\dg\hat d_{k}}
\\ \avg{\hat d_{q}\hat d_{p+q}\dg\hat h_{p-k}\hat d_{k}} & \approx \gd_{p,0}\avg{\hat d_{q}\hat d_{q}\dg}\avg{\hat d_{k}\hat h_{-k}} + \gd_{p+q,k}\avg{\hat d_{q}\hat h_{-q}}\avg{\hat d_{k}\dg\hat d_{k}} - \tfrac{1}{N_{s}}\tfrac{1}{n_{d}}\avg{\hat d_{q}\hat h_{-q}}\avg{\hat d_{p+q}\dg\hat d_{p+q}}\avg{\hat d_{k}\dg\hat d_{k}}.
\end{split} \end{equation}

Applying these approximations to \cref{eq:eomfull}, we find
\begin{equation} \begin{split}
\frac{d}{dt}\avg{\hat d_{k}\dg\hat d_{k}} & = i J \tilde n\gve_{k}\p{\avg{\hat d_{k}\hat h_{-k}} - \avg{\hat d_{k}\dg\hat h_{-k}\dg}}
 -iJ\tilde n\br{\begin{array}{c}
3\p{\gve_{k}n_{d}\avg{\hat d_{k}\hat h_{-k}} - \gve_{0}\eta\avg{\hat d_{k}\dg\hat d_{k}}} 
\\ + 2\gve_{0}\p{\sqrt{1 + \tfrac{1}{4\tilde n^{2}}}\eta^{*}\avg{\hat d_{k}\hat h_{-k}} + \xi_{d}\avg{\hat d_{k}\hat h_{-k}} }
\end{array} - \hc}
\\ \frac{d}{dt}\avg{\hat d_{k}\hat h_{-k}} & = -i U\avg{\hat d_{k}\hat h_{-k}}
 - iJ\tilde n\gve_{k}\p{1  - 6n_{d} + 9\p{n_{d}^{2} - \xi_{d}^{2}} + 6\abs{\eta}^{2}}
\\ & \quad -2iJ\tilde n \br{\sqrt{1 + \tfrac{1}{4\tilde n^{2}}}\p{\avg{\hat d_{q}\hat h_{-q}} - \eta} + \avg{\hat d_{k}\dg\hat d_{k}} - \xi_{d}}
\\ & \quad + 2i J \tilde n \bmat{
\sqrt{1 + \tfrac{1}{4\tilde n^{2}}}\mat{3\p{\gve_{k}n_{d}\avg{\hat d_{k}\hat h_{-k}} - \gve_{0}\eta\avg{\hat d_{k}\dg\hat d_{k}}} +  2\gve_{0}\xi_{d}\avg{\hat d_{k}\hat h_{-k}}}
\\ + 3\p{\gve_{k}n_{d}  - \gve_{0}\xi_{d}}\avg{\hat d_{k}\dg\hat d_{k}} + 3\gve_{0}\eta^{*}\avg{\hat d_{k}\hat h_{-k}} 
}
\label{eq:eomapx}
\end{split} \end{equation}
with $\avg{\hat h_{-k}\dg\hat h_{-k}} = \avg{\hat d_{k}\dg\hat d_{k}}$.

\bibliographystyle{apsrev}
\bibliography{/Users/yarivyanay/Documents/University/Citations/library}

\end{document}